\documentclass[prd,twocolumn,a4paper,superscriptaddress,floatfix,showpacs]{revtex4}
\usepackage{graphicx}
\usepackage{amssymb}
\usepackage{amsmath}
\newcommand{\be}{\begin{equation}}
\newcommand{\ee}{\end{equation}}
\newcommand{\ben}{\begin{eqnarray}}
\newcommand{\een}{\end{eqnarray}}
\newcommand{\bes}{\begin{subequations}}
\newcommand{\ees}{\end{subequations}}
\newcommand{\bb}{\bibitem}

\begin{document}
\title{Impact of Lorentz violation on the dynamics of inflation}
\author{P.P. Avelino}
\affiliation{Centro de F\'{\i}sica do Porto and Departamento de F\'\i sica, Faculdade de Ci\^encias, Universidade do Porto\\
Rua do Campo Alegre 687, 4169-007 Porto, Portugal}
\author{D. Bazeia}
\affiliation{Departamento de F\'{\i}sica, Universidade Federal da Para\'{\i}ba
58051-900 Jo\~ao Pessoa, Para\'{\i}ba, Brasil}
\author{L. Losano}
\affiliation{Departamento de F\'{\i}sica, Universidade Federal da Para\'{\i}ba
58051-900 Jo\~ao Pessoa, Para\'{\i}ba, Brasil}
\author{R. Menezes}
\affiliation{Centro de F\'{\i}sica do Porto and Departamento de F\'\i sica, Faculdade de Ci\^encias, Universidade do Porto\\
Rua do Campo Alegre 687, 4169-007 Porto, Portugal}
\affiliation{Departamento de F\'{\i}sica, Universidade Federal da Para\'{\i}ba 58051-900 Jo\~ao Pessoa, Para\'{\i}ba, Brasil}
\author{J.J. Rodrigues}
\affiliation{Departamento de F\'{\i}sica, Universidade Federal da Para\'{\i}ba 58051-900 Jo\~ao Pessoa, Para\'{\i}ba, Brasil}
\date{\today}

\begin{abstract}
This work deals with the dynamics of inflation in the context of a scalar-vector-tensor theory of gravity exhibiting spontaneous Lorentz 
violation at early times. We describe a first-order formalism which we use to obtain new exact Lorentz violating inflationary solutions for a broad family of models, some in the absence of a potential for the inflaton field. Our results show that different conditions are required to solve the horizon and flatness problems. In particular, we find a necessary condition for inflation to provide a solution to both problems and we show that in inflationary models with no inflaton potential a period of superinflation might be necessary to solve the flatness problem.
\end{abstract}


\maketitle
\section{Introduction}

The Einstein-Aether Theory is a natural framework to explore the violation of the Lorentz symmetry in a cosmological context, where gravity plays a dominant role. This theory has a vector field which couples non-minimally to the metric tensor. It is assumed that its norm has spontaneously acquired a non-zero expectation value thus defining a preferred frame, leading to a spontaneous violation of local Lorentz invariance. This was first investigated by Will and Nordtvedt in 70's \cite{WN1} and recently studied in detail by  Jacobson and Mattingly \cite{JM1}, based on ideas of Kostelecky and Samuel \cite{KS1}. 

The parameter space of theory, at the present time,  is severely restricted both by Parametrized Post Newtonian \cite{FJ} and pulsar constraints \cite{F}. On the other hand, the observation of ultra-high energy cosmic rays implies no significant energy loss via vacuum Cherenkov type radiation thus leading to very stringent constraints on the parameters of theory \cite{EMS}. Still,  cosmological constraints on the violation of the Lorentz symmetry, in the context of the Einstein-Aether theory, have been investigated by various authors which found that the theory is consistent with experimental and observational data over a range of its parameter space. On small scales the Einstein-Aether vector field will in general lead to a renormalization of the local Newton Constant \cite{CL} while on large scales it may leave an observable imprint on the galaxy and angular CMB power spectra 
\cite{L,LMB,ZFZ} (see also \cite{AZTG,ZET}).

Other related investigations concerning the breaking of the Lorentz symmetry in cosmology  due to a vector field include vector-like dark energy \cite{r1}, Lorentz violating brane world scenarios \cite{r2} and Baryogenesis via spontaneous Lorentz violation \cite{r4}.

In this paper we shall investigate a simple extension of the Einstein-Aether theory in the context of inflation which includes 
a scalar field non-minimally coupled to gravity as an extra ingredient. We shall assume that this scalar field is the inflaton field, 
driving a period of accelerated expansion in the early universe. We use a simple but powerful first-order method introduced in ref. \cite{bglm} to obtain exact analytical solutions (see also \cite{blrr}).

Among the results to be obtained below, we shall extend approximate results obtained in \cite{KS2} and obtain some new exact solutions for several distinct models. In this context we will find a necessary condition which has to be satisfied if inflation is to provide a solution to the flatness problem. Particular attention will be given to models with no inflaton potential, where inflation is fully associated with the Lorentz violation. Throughout the work, we will assume the signature $[-,+,+,+]$ and use units in which $4\pi G=1$.

\section{The Einstein-Aether Theory}

The Einstein-Aether Theory is described by the action
\be
S=\int{d^{4}x\sqrt{-g}\left(\frac{R}{4}+{\cal L}\right)}\,, \label{action}
\ee
where $R$ is the Ricci scalar, ${\cal L}={\cal L}_m+{\cal L}_V$, ${\cal L}_m$ is the standard matter Lagrangian and ${\cal L}_V$ is given by
\be
{\cal L}_V={K^{\mu \nu}}_{\alpha \beta} \nabla_\mu A^\alpha \nabla_\nu A^\beta + \lambda (A^\mu A_\mu+1)\,,
\ee
with 
\be
{K^{\mu \nu}}_{\alpha \beta} = - c_1 g^{\mu \nu} g_{\alpha \beta} - c_2 \delta^\mu_\alpha \delta^\nu_\beta - c_3 \delta^\mu_\beta \delta^\nu_\alpha - c_4 A^\mu A^\nu g_{\alpha \beta}\,.
\ee
Here $g^{\mu \nu}$ and $A^\mu$ are respectively the components of the metric tensor and Einstein-Aether vector field. General Relativity is recovered for $c_1=c_2=c_3=c_4=0$ \cite{AZTG}. 

Although the action (\ref{action}) is Lorentz invariant we shall assume that Lorentz invariance is spontaneously broken with the norm of the Einstein-Aether vector field acquiring a non-zero expectation value, $A_\mu A^\mu=-1$. The requirement that Einstein-Aether vector is timelike is necessary for the stability of the theory \cite{EMS,r3}. 

In this paper we shall consider a more general Scalar-Vector-Tensor theory \cite{KS2} ${\cal L}={\cal L}_m+{\cal L}_V + {\cal L}_\phi$, with the coefficients $c_1$, $c_2,$ $c_3$ and $c_4$ now being functions of a real scalar field $\phi$ which is described by 
\be
{\cal L}_\phi=-\frac12\nabla_\mu\phi\nabla^\mu\phi-V(\phi)\,,
\ee
where $V(\phi)$ is the scalar field potential.  

\section{Equations of motion}

We shall consider a homogeneous and isotropic Friedmann-Robertson-Walker universe with line element
\be
ds^2=-dt^2+a^2(t)\left(\frac{dr^2}{1-kr^2}+r^2\left(d\theta^2+\sin^2 \theta d \varphi^2\right)\right)\,,
\ee
where $t$ is the physical time, $(r,\theta,\varphi)$ are spatial comoving coordinates, $a(t)$ is the scale factor and $k$ is the constant spatial curvature. The Einstein-Aether vector field has to be homogeneous and isotropic and consequently $A^{\mu}=\left(1,0,0,0\right)$.
If we take ${\cal L}_m=0$ then the dynamics of the universe is fully described by the following equations of motion
\ben
(1+2\beta)H^2=\frac23\left(\frac12\dot{\phi}^2+V\right)-\frac{k}{a^2}\,,\label{h2}
\\
(1+2\beta)\dot{H}=-\dot{\phi}^2-2\beta_{\phi}H\dot{\phi}+\frac{k}{a^2}\,,\label{hdot}
\een
where $\beta(\phi)=c_1(\phi)+3 c_2(\phi)+c_3(\phi)$. In this paper we shall assume that $1+2 \beta > 0$.

Note that the contribution of the curvature term in the right hand side of Eq.~\eqref{h2} does not necessarily decrease during inflation (defined as an epoch in which $\ddot a>0$). The critical density is given by
\be
\rho_c=\frac32 (1+2\beta) H^2\,,
\ee
so that
\be
\Omega -1= \frac{3k}{2a^2 \rho_c}= \frac{k}{a^2 (1+2\beta) H^2}\,,
\ee
where $\Omega \equiv \rho/\rho_c$.
If $|\Omega-1|$ is a decreasing function of physical time then
\be\label{flat}
\frac{d(H^2 a^2(1+2\beta))}{dt}>0\,,
\ee
rather than the inflationary condition $d(H^2 a^2)/dt>0$. The acceleration parameter is defined as
\be
\label{q}
q \equiv \frac{\ddot a}{aH^2}=\frac{1}{2 H^3 a^2} \frac{d}{dt}\left(a^2 H^2\right)\,.
\ee
If $q <0$ we have deceleration and consequently $Ha$ decreases with cosmic time, $t$. On the other hand $q>0$ implies that the universe is accelerating (inflating) and, in that case, $Ha$ is a growing function of $t$. For $q = 1$ the Hubble parameter is constant and the expansion is exponential. If $q>1$ the Hubble parameter grows with $t$ and we have superinflation. 

We may also define an analogous parameter
\be
\label{qf}
q_k \equiv \frac{1}{2 H^3 a^2 (1+2\beta)} \frac{d}{dt}\left(a^2 H^2 (1+2\beta)\right)\,,
\ee
which is greater (smaller) than zero depending on whether $\Omega=1$ is an attractor (repulsor) -- note that $q_k = q$ if $\beta$ 
is a constant but in general $q_k \neq q$. Hence, one may have an inflationary regime that solves the homogeneous but not the flatness problem. From now on we shall assume that the universe is approximately flat during most of inflationary regime. Consequently, we shall take $k=0$ in the determination of the evolution of $a$ and $\phi$ with physical time. 

The parameter $q_k$ can also be written as 
\be
q_k=q+\frac{1}{H}\frac{d}{dt}\left(\ln(1+2\beta)\right)\,,
\ee
and we see that $q_k > q$ or $q_k < q$ depending on whether $d \beta/dt$ is greater or smaller than zero, respectively.

In this paper we will neglect the radiation and matter content of the universe in the inflationary regime. 
If inflation is to solve the flatness problem, this is expected to be a good approximation during most of the inflationary era since the ratio between the scalar field energy density
\be
\rho_\phi=\frac12{\dot\phi}^2+V(\phi)\,,
\ee
and the matter and radiation energy densities must grow considerably during inflation.

The equation of motion for the scalar field, $\phi$, can also be written explicitly as
\be\label{EC}
\ddot{\phi}+3H\dot{\phi}+3H^2\beta_{\phi}+V_{\phi}=0\,.
\ee
If $\beta=0$ we recover the dynamics of the standard cosmology. 

\section{First-order formalism}

Here we shall use the first-order formalism introduced in ref. \cite{bglm}. This is important to investigate examples from which we can obtain exact analytical solutions. The starting point is to assume $\dot \phi$ is a function of $\phi$ alone,
\be\label{FO}
\dot{\phi}=-W_{\phi}\,,
\ee
so that
\be\label{FO1}
t-t_0=-\int_{\phi_0}^{\phi} \frac{d\phi}{W_\phi}\,,
\ee
where $W(\phi)$ is an arbitrary function of $\phi$. Note that Eq.~\eqref{FO} is a first-order equation while  Eq.~\eqref{EC} is second-order.

Substituting Eq.~\eqref{FO} in Eq.~\eqref{hdot} one obtains (taking $k=0$)
\be\label{hdot1}
(1+2\beta)\frac{dH}{d\phi}=W_\phi-2\beta_{\phi}H\,,
\ee
or equivalently
\be\label{hdot2}
H(\phi)=\frac{W(\phi)+C}{1+2\beta(\phi)}\,,
\ee
where $C$ is an arbitrary constant. In the following we shall take $C=0$ so that
\be
H(\phi)=\frac{W(\phi)}{1+2\beta(\phi)}\,.\label{hphi}
\ee
There is a family of solutions for $H(\phi)$ (and $W(\phi)$) which correspond to the same $\beta(\phi)$ and scalar field potential, $V(\phi)$. This is a result of the freedom to fix the kinetic energy of the scalar field at a given initial time. We are also implicitly assuming that there is only one value of $H$ corresponding to each value of $\phi$. This assumption is valid in many situations of cosmological interest but may not be 
true in general (for example if $\phi$ is oscillating around a minimum of the potential). However, it will be satisfied during the inflationary regime we will be dealing with in the present work.

The expansion factor can now be computed using Eqs.~\eqref{FO} and \eqref{hphi}  as
\be
\label{sfac}
a(\phi)=a_0 e^{-\int_{\phi_0}^\phi \frac{W}{W_\phi} \frac{d\phi}{1+2\beta}}\,.
\ee
The subscript `0' denotes that the physical quantities are to be evaluated at some initial time $t_0$. 
This can be used to calculate the number of e-foldings of inflation, $N$, between $\phi_0$ and $\phi_e$ 
\be
N={-\int_{\phi_0}^{\phi_e} \frac{W}{W_\phi} \frac{d\phi}{1+2\beta}}=\ln\left(\frac{a(\phi_e)}{a_0}\right)\,,
\ee
where the subscript `e' refers to the end of inflation defined by ${\ddot a} (t_e) =0$. 
The acceleration parameter, $q$, is given by
\ben\label{acel}
q &\equiv& \frac{\ddot a}{a H^2}=\frac{\dot H}{H^2}+1 = 
-\frac{d}{dt}\left(\frac{1}{H}\right)+1=\nonumber
\\
& &\frac{d}{d\phi}\left(\frac{1+2\beta}{W}\right)W_\phi+1\,,
\een
and is greater than zero during inflation. The value of $\phi$ at the end of the inflationary regime ($\phi_e$) can be calculated 
by solving the equation $q(\phi)=0$. The parameter $q_k$ is equal to
\be
\label{qf1}
q_k=q-\frac{\beta_\phi W_\phi}{W}\,.
\ee

Using Eq.~\eqref{h2} we can now obtain the corresponding potential for the scalar field
\be\label{V}
V(\phi)=\frac32\frac{W^2}{1+2\beta}-\frac12W_{\phi}^2\,,
\ee
which represents an extension for $\beta \neq 0$ of the result given in \cite{bglm,SB,ST}.

A special case in which Eq.~\eqref{hphi} applies is that of a slow roll stage satisfying ${\dot{\phi}}^2 \ll V$ so that 
\be
V(\phi) \sim \frac32\frac{W^2}{1+2\beta}\,.
\ee

\section{Analytical solutions}

We will now apply the first-order formalism described in the previous section to obtain new exact Lorentz violating inflationary solutions for a broad family of models. The simplest choice is to make $\beta$ a constant, but this leads to $q_k=q$. Other more interesting possibilities are obtained if $\beta$ has an explicit dependence on the scalar field, $\beta=\beta(\phi)$. Some particular cases of current interest will be considered below. 

\subsection{Exponential expansion ($q=1$)}

Let us start with the special case of exponential expansion which requires that the Hubble parameter, $H$, is a constant ($q=1$). In this case the scale factor, $a$, grows exponentially with physical time ($a \propto \exp(Ht)$). If $\beta=0$ then $H=W={\rm constant}$ and $\dot{\phi}=- W_\phi = 0$ so that $V=3H^2/2 = {\rm constant}$. On the other hand, if $H={\rm constant}$ and $\beta \neq 0$ then 
\be
V = \frac32 H^2 (1+2 \beta) -2 H^2 \beta_\phi^2\,,
\ee
which is no longer required to be a constant. In this case $\dot{\phi}=-W_\phi=-2H \beta_\phi$ so that
\be
\int_{\phi_0}^{\phi} \frac{d\phi}{\beta_\phi} =- 2 H (t-t_0)\,.
\ee
If $V=0$ then 
\be
1+2\beta=\frac14\left({\sqrt 3} \left(\phi-\phi_0\right)+2(1+2\beta_0)^{1/2}\right)^2\,,
\ee
for exponential expansion to occur.  Note that in this solution $\beta \to -1/2$ when $\phi \to \phi_0-2 {\sqrt 3} (1+2\beta_0)^{1/2}/3$. 
Hence, this solution can only be valid in a Lorentz violating period of inflation and does not account for the transition to a standard slow roll stage with $|\beta| \ll 1$. We will discuss this transition below when we consider a polinomial solution for $W$ and $\beta$.

The evolution of $\phi$ with physical time is given by
\ben
\left(\frac{3}{4}(\phi-\phi_0)+\frac{\sqrt{3}}{2}(1+2\beta_0)^{1/2}\right)^{2/3}=\nonumber
\\
\left(\frac{\sqrt3}{2}(1+2\beta_0)^{1/2}\right)^{2/3}e^{-H(t-t_0)}\,.
\een

\subsection{The null potential}

An interesting case occurs when the scalar field potential is null ($V=0$) and we can write
\be
\label{WV0}
\frac{W_{\phi}^2}{W^2}=\frac{3}{1+2\beta}\,,
\ee  
so that 
\be
W(\phi)=W_0 e^{{\sqrt 3}\int_{\phi_0}^\phi (1+2\beta)^{-1/2}}d\phi\,.
\ee
In this case, the Hubble parameter is $H=W_\phi^2/(3W)$ and consequently, using Eq.~\eqref{sfac}, we 
obtain
\be
a=a_0 \left(\frac{W}{W_0}\right)^{-1/3}\,.
\ee

For simplicity we take 
\be
W(\phi)=W_0\left(\frac{\phi}{\phi_0}\right)^n \label{wphi}\,,
\ee
with $n > 0$ and $W_0>0$. Using Eq. (\ref{FO1}) we obtain the evolution of $\phi$ with physical time
\ben
\phi&\!=\!&\phi_0\!\left(\!\frac{n(n\!-\!2)W_0}{\phi_0^2}(t-t_0)+1\!\right)^{1/(2-n)}\!\!\!, \ \  n \neq 2\,, \label{pol1}
\\
\phi&=&\phi_0 \, e^{-2 W_0(t-t_0)/\phi_0^2}\,, \ \  n = 2 \,. \label{pol2}
\een 
Note that if $n \ge 2$ then $\phi \to 0$ when $t \to +\infty$ while if $n <2$ then $\phi \to 0$ when $t \to  \phi_0^2/(n(2-n)W_0)$.

Eqs.~\eqref{WV0} and \eqref{wphi} imply
\be
1+2\beta=\frac{3\phi^2}{n^2}\,.
\ee 
In this case we have
\ben
H=\frac{n^2 W_0}{3 \phi_0^2}\left(\frac{\phi}{\phi_0}\right)^{n-2}\,,
\\
a=a_0 \left(\frac{\phi}{\phi_0}\right)^{-n/3}\,,
\een
where the evolution of $\phi$ with physical time is given by Eqs.~\eqref{pol1} and  \eqref{pol2}.

Inflation is defined as an epoch satisfying $\ddot a>0$ or equivalently as
\be
{\dot a}=aH\propto\phi^{2n/3-2}\,,
\ee
is a growing function of time, which happens for $n < 3$.  If $n=2$ then $H$ is constant and the scalar factor grows exponentially with physical time, $a=a_0\exp[H(t-t_0)]$. If $n<2$ then $H$ grows with time which corresponds to superinflation. In this case, we have that $a\to\infty$ when $t\to t_0+\phi_0^2/(n(2-n)W_0)$.

As discussed before, the inflationary condition is not sufficient to 
guarantee that the curvature term in Eq.~\eqref{h2} becomes less important for increasing physical times. Such condition, 
necessary if the model is to provide a solution to the flatness problem, requires that  Eq.~\eqref{EC} is satisfied or equivalently 
that
\be
a^2H^2 (1+2\beta) \propto \phi^{4n/3-2}\,,
\ee
is a growing function of time. This happens for $n<3/2$. Hence, for $3/2<n<3$ inflation occurs but does not solve the flatness 
problem independently of how long it takes. This can also be seen calculating $q$ and $q_k$ using Eqs.~\eqref{acel} and 
\eqref{qf1} respectively
\ben
q &=& \frac{3(2-n)}{n}+1\,,
\\
q_k &=& q- \frac{3}{n}= \frac{3(1-n)}{n}+1\,.
\een
If $q <0$ ($n>3$) we have deceleration while $q>0$  ($n<3$) implies that the universe is accelerating (inflating). For $q = 1$ ($n=2$) the Hubble parameter is constant and the expansion is exponential. If $q>1$ ($n<2$) the Hubble parameter grows with physical time and we have superinflation. In order for inflation to solve the flatness problem the condition $q_k>0$ needs 
to be satisfied which happens for $n<3/2$. Hence, in this model the solution of the flatness problem requires a period o superinflation. 

\subsection{The polynomial solution}

Taking $W(\phi)$ as in Eq.~\eqref{wphi}, the corresponding potential is 
\be
V(\phi)=\frac12 W_0^2 \left(\frac{\phi}{\phi_0}\right)^{2n}\left(\frac{3}{1+2\beta(\phi)}-\frac{n^2}{\phi^2}\right)\,,
\ee
which depends on $\beta(\phi)$ (note that the evolution of $\phi$ with physical time is again given by Eqs.~\eqref{pol1} and \eqref{pol2}). If we now choose $\beta(\phi)$ as a quadratic function of the scalar field, in the form
\be
\beta(\phi)=\beta_0 \left(\frac{\phi}{\phi_0}\right)^2\,,
\ee
then the scale factor is given by
\be
a(\phi)=a_0 \left(\frac{1+2\beta_0(\phi/\phi_0)^2}{1+2\beta_0}\right)^{-\phi_0^2/(4 n \beta_0)}\,.
\ee
In this case $\beta\to0$ when $\phi\to0$ which means that this solution accounts for the transition to a phase where 
the coupling to the Lorentz violating sector becomes negligible.

The parameters $q$ and $q_k$ are 
\ben
q &=& -\frac{n^2}{\phi^2} +\frac{2n(2-n)\beta_0}{\phi_0^2}+1\,,
\\
q_k &=& q - \frac{2n \beta_0}{\phi_0^2}=-\frac{n^2}{\phi^2} +\frac{2n(1-n)\beta_0}{\phi_0^2}+1\,,
\een 
and the value of $\phi$ at the end of inflation, satisfying $q(\phi_e)=0$, is given by
\be
\phi_e=n\left(2n(2-n)\frac{\beta_0}{\phi_0^2}+1\right)^{-1/2}\,.
\ee
Here we are implicitly assuming that $\phi>0$ and we see that inflation can only take place if $2n(2-n)\beta_0/ \phi_0^2 +1 > 0$.
On the other hand, the condition $q_k(\phi)=0$ is satisfied for
\be
\phi_c=n\left(2n(1-n)\frac{\beta_0}{\phi_0^2}+1\right)^{-1/2}\,,
\ee
assuming that $2n(1-n)\beta_0/ \phi_0^2 +1 > 0$.
We see that if $\beta_0<0$, then $\phi_c>\phi_e$ is reached at physical times $t_c<t_e$; however, if $\beta_0>0$, then $\phi_c<\phi_e$ is reached at physical times $t_c>t_e$. Consequently, if $\beta_0 >0$ then $\Omega$ deviates from unity in the last stage of inflation (for $\phi_e<\phi <\phi_c$). On the other hand, $\Omega$ may approach unity, even without inflation, if $\beta_0 < 0$. 

Here we also calculate the number of e-foldings of inflation, which is given by
\be
N=-\frac{\phi_0^2}{4n\beta_0}\ln\left(\frac{\phi_0^2+4n\beta_0}{(1+2\beta_0)(\phi_0^2-2n(n-2)\beta_0)}\right)\,.
\ee

We may alternatively choose to take
\be
\label{betadef}
1+2\beta(\phi)=\left(1+2\beta_0\right) \left(\frac{\phi}{\phi_0}\right)^m\,,
\ee
with $1+2\beta_0>0$ and $m>0$ (note that the contribution to the Lorentz violation vanishes for $\beta_0=m=0$). If  $\beta_0 \neq 0$ or $m \neq 0$ then $\beta$ does not vanish at the end of inflation ($\beta\to-1/2$ when $\phi\to0$). In this scenario the coupling to the Lorentz violating sector does not switch off at the end of inflation and consequently it may not account for the whole evolution of the scalar field in the inflationary regime. Still, Eq. (\ref{betadef}) allow us to increase the model complexity, by adding one extra parameter, while retaining the ability to find analytical solutions for the universe dynamics.

With the above choice for $\beta(\phi)$ we can write 
\be
a=a_0 \left(\frac{\phi}{\phi_0}\right)^{-\frac{\phi_0^2}{n(1+2\beta_0)}}\,, 
\ee
for $m=2$ and
\be
a=a_0\exp\left(\frac{\phi_0^2}{n(2-m)(1+2\beta_0)}\left(1-\left(\frac{\phi}{\phi_0}\right)^{2-m}\right)\right)\,, 
\ee
if $m \neq 2$. Note that $a$ will tend to a constant or to $\infty$ depending on whether $m$ is smaller or greater than zero, 
respectively. The parameters $q$ and $q_k$ are given by
\ben
q&=&\frac{n(m-n)(1+2\beta_0)}{\phi_0^2}\left(\frac{\phi}{\phi_0}\right)^{m-2}+1\,,
\\
q_k &=& q - \frac{mn (1+2\beta_0)}{2\phi_0^2}\left(\frac{\phi}{\phi_0}\right)^{m-2}\,,\nonumber
\\
&=&\frac{n(m-2n)(1+2\beta_0)}{2\phi_0^2}\left(\frac{\phi}{\phi_0}\right)^{m-2}+1\,.
\een
We see that if $m \ge n$ or $m \ge 2$ then inflation cannot come to an end (if $q > 0$ at $t=t_0$ then it never becomes 
negative at $t>t_0$). In particular, if $m=2$ then both $q$ and $q_k$ are constants. On the other hand, if $m=n$ then $q=1$, leading to an exponential expansion. Also note that, for $n<2$ and $m >n$, the inflationary stage never stops despite the fact that the scalar factor, $a$, is approaching a constant as $t \to \phi_0^2/(n(2-n)W_0)$.

Hence, any interesting solutions, with a transition from an accelerating to a decelerating phase, have $m<n$ and $m<2$. In this case, at the end of inflation, for $q(\phi_e)=0,$ we get, 
\be
\phi_e=\phi_0\left(\frac{\phi_0^2}{n(n-m)(1+2\beta_0)}\right)^{1/(m-2)}\,,
\ee
and the condition $q_k(\phi_c)=0$ gives 
\be
\phi_c=\phi_0\left(\frac{2\phi_0^2}{n(2n-m)(1+2\beta_0)}\right)^{1/(m-2)}\,,
\ee
with $\phi_c < \phi_e$. Again, we see that $\Omega$ may deviate from unity during the last stage of inflation. 
\section{Conclusion}

We have performed a detailed study of the impact of Lorentz violation on the dynamics of inflation, considering an extension of the 
Einstein-Aether theory. We have obtained a broad class of exact Lorentz violating inflationary solutions for the evolution of 
the scale factor and the inflaton field, some in the absence of a potential for the inflaton field, using a first-order formalism similar to that used in \cite{bglm}. The presence of first-order differential equations is crucial for the analytical results, and it has helped us to construct a diversity of exactly solved examples of current interest to inflationary cosmology. We have shown that, unlike in the standard inflationary scenario, in the context of Lorentz violating inflationary scenarios different conditions are required in order  to solve the horizon and flatness problems and we found a necessary condition for inflation to provide a solution to both problems. In particular, in Lorentz invariant inflationary models with no inflaton potential a period of superinflation was shown to be required in order to solve the flatness problem.

This work is supported in part by the CAPES-GRICES joint project between DF-UFPB, Brazil, and DF-UP, Portugal. The authors also thank CNPq, PRONEX-CNPq-FAPESQ, and FCT for partial support. R.M. thanks CAPES for the fellowship BEX-0299/08-1.


\end{document}